\documentclass[a4paper,12pt]{article}
\usepackage{amsmath,amsfonts}
\def\ba {\begin {array}}
\def\ea{\end {array}}
\def\be {\begin {equation}}
\def\ee {\end {equation}}
\def\bea {\begin {eqnarray}}
\def\eea {\end {eqnarray}}

\newcommand{\bi}{\begin{itemize}}
\newcommand{\ei}{\end{itemize}}

\newcommand{\bb}[1]{\makebox[16pt]{{\bf#1}}}

\newtheorem{theorem}{Theorem}

\newtheorem{lemma}{Lemma}

\newtheorem{corollary}{Corollary}
\newtheorem{example}{Example}

\begin{document}
\title{Laplace-type  equations as conformal superintegrable systems}
\author{E. G.~Kalnins\\
Department of Mathematics,\
University
of Waikato,\\
 Hamilton, New Zealand.\\
J.~M.~Kress\\ 
 School of Mathematics, The University of New South Wales, \\
Sydney NSW 2052, Australia \\
W.~Miller, Jr.\\
 School of Mathematics, University of Minnesota,\\
Minneapolis, Minnesota, U.S.A.\\
S.~Post\\
 School of Mathematics, University of Minnesota,\\
Minneapolis, Minnesota, U.S.A.}

\maketitle

\begin{abstract} We lay out the  foundations of the theory of second-order  conformal superintegrable systems.  Such systems are essentially Laplace equations on a manifold with an added potential: $(\Delta_n+V({\bf x}))\Psi=0$. Distinct families of second-order superintegrable Schr\"odinger (or Helmholtz) systems $(\Delta'_n+V'({\bf x}))\Psi=E\Psi$ can be incorporated into a single Laplace equation. There is a deep connection between most of the special functions of mathematical physics, these Laplace conformally superintegrable systems and their conformal symmetry algebras. Using the theory of the Laplace systems, we show that the problem of classifying all 3D Helmholtz superintegrable systems with nondegenerate potentials, i.e., potentials with a maximal number of independent paprameters,  can be reduced to the problem of classifying the orbits of the nonlinear action of the conformal group on a 10-dimensional manifold. 
\end{abstract}

\thanks{Dedicated to our colleague Dennis Stanton}


\section{Introduction}
From our point of view special functions are, in large part,   ``special" because they  arise from mathematical models of  physical systems that are completely solvable analytically and algebraically. Intuitively we consider such systems to be of high symmetry, but this symmetry may be ``hidden", i.e., not obvious.  Special function theory can be based on the notion of superintegrability; it is the best concept to date to capture both hidden symmetry and just those systems whose associated functions are interesting and useful enough to be considered ``special". An $n$-dimensional Hamiltonian system (2n-dimensional phase space), classical or quantum, is integrable if it admits n functionally independent commuting constants of the motion (or symmetry operators), polynomial in the momenta. It is superintegrable if it is integrable and admits 2n-1 constants of the motion (the maximum possible but, of course, not all commuting). If  the functionally independent constants of the motion can all be chosen of order k or less in the momenta (or in the derivatives for the quantum case) the system is called kth order superintegrable. Superintegrability is a much stronger requirement than integrability, indeed superintegrable systems can be solved algebraically.

Special functions are connected to superintegrable systems in several ways. The most obvious is that they occur when one computes the eigenfunctions of the Hamiltonian operator or other symmetry operator in a 1st or 2nd order quantum superintegrable system. The superintegrability forces variable separability, usually in multiple systems. The  majority of special functions of mathematical physics arise from separable coordinates in this way. The symmetry algebras generated by 1st order quantum superintegrable systems are  Lie algebras and  the typical special functions arising are spherical harmonics, and other  orthogonal polynomials \cite{Vilenkin, Stanton}.  Superintegrable systems of 2nd order are multi-integrable (so multi-separable) and  distinct classes of special functions can be related to one another in a single system. Indeed, the basic properties of Gaussian hypergeometric functions and their various limiting cases, as well as Lam\'e, Mathieu and Heun functions, and ellipsoidal harmonics are associated with 2nd order superintegrable quantum systems via separation of variables. The algebra formed by the  generating   symmetries  again closes under commutation.  For example, consider  operator superintegrable systems of the form $(\Delta +V)\Psi=E \Psi$ on a 2-dimensional conformally flat manifold with potential function $V$,  and symmetry generators of order no more than 2. Then the system of symmetries closes, sometimes at order 1 (Lie algebras and Lie groups, $V=0$), sometimes at order 3  for degenerate potentials, and sometimes at order 6 for non-degenerate  potentials. There are no other possibilities. Each such system is multiseparable. (For $n>2$ similar statements hold but  there are more possibilities). Closure at order 1 corresponds to a Lie algebra.  The monograph \cite{Miller77}, written before the word ``superintegrable'' was coined, is really about some simple 2nd order superintegrable systems whose algebra closes at order 1. 

Closure at orders 3 or 6 defines quadratic algebras (NOT Lie algebras) whose algebraic representation theory gives crucial information about the possible energy eigenvalues $E$ and the expansion of one integrable eigenbasis in terms of another (i.e., the expansion of one class of special functions in terms of another). The representation theory of these quadratic algebras is of great intrinsic interest and leads to another connection with the theory of special functions.  For example, consider the system $(\Delta +a/(s_1)^2+b/(s_2)^2+c/(s_3)^2)\Psi=E \Psi$  where  $\Delta$ is the Laplace-Beltrami operator on the 2-sphere $(s_1)^2+(s_2)^2+(s_3)^2=1$. This is 2nd order superintegrable with a quadratic algebra that closes at order 6 (i.e., differential operators of order 6). It is an amazing fact that for eigenvalues $E$ of finite multiplicity, this algebra is precisely  the structure algebra for the Racah polynomials in their full generality. The algebra for the infinite dimensional bounded below representations of the quadratic algebra yields the Wilson polynomials in their full generality \cite{KMPost, AAR, KMPost4, Post2009}. Thus all of the classical discrete orthogonal  polynomials and their Wilson polynomial generalizations appear naturally in the representation theory of the quadratic algebra. The interpretation here is that the Schr\"odinger equation separates in spherical coordinates and elliptic coordinates. The spherical coordinate eigenfunctions are products of Jacobi polynomials, known as Karlin-McGregor 2-variable orthogonal polynomials in this case. The elliptic coordinate eigenfunctions are products of Heun polynomials. The expansion of the 
Heun polynomial basis in terms of the  Karlin-McGregor basis can be computed directly from the representation theory of the quadratic algebra, but is complicated. One can also choose spherical coordinates based on rotation about the 3-axis or spherical coordinates based on rotation about the 2-axis. The  coefficients giving the expansion of one spherical coordinate basis in terms of the other are just the Racah (or Wilson) polynomials. Thus, study of the representation theory of the quadratic algebra leads to classes of special functions, such as orthogonal polynomials of a discrete variable, that do not arise directly from variable separation.

Special functions are also associated with higher order superintegable systems. Thus the  Painlev\'e transcendents (not associated with variable separability) appear in the study of 3rd order superintegrable systems and the representations of their cubic symmetry algebras, \cite{GW, Gravel,IM2009}. Similarly, superintegrability can be related to  generalized hypergeometric functions and many variable hypergeometric functions (Appell functions, Lauricella functions, Horn functions, etc.)  For 1st order superintegrability some of this is in \cite{Miller77}, but this is mostly virgin territory for superintegrability of order $\ge 2$.

Superintegrabilty can also  be studied for equations of the form $(\Delta_n +V)\Psi=0$ on  conformally flat $n$-dimensional manifolds, and that is the subject of this paper. For example, Volkmer \cite{Volkmer2006} treated the equation  $(\Delta_n +a_1/(x_1)^2+...+a_n/(x_n)^2)\Psi=0 $ on $ n$-dimensional flat space, in a study of generalizations of ellipsoidal wave functions. Such functions arise when this equation is separated in conical coordinates. From our point of view, this is a conformally 2nd order superintegrable system with degenerate potential. The added insight provided by superintegrability theory is that the  conformal symmetry algebra of this equation closes at order 6. Thus the representation theory of this quadratic algebra will provide important spectral information about the symmetry operators of the Laplace equation.

This paper  inaugurates the study of  second-order  conformally superintegrable operator systems on conformally flat manifolds, essentially Laplace equations on a manifold with an added (very special) potential: $(\Delta_n+V({\bf x}))\Psi=0$. There are several important features of this approach. First, distinct families of second-order superintegrable Schr\"odinger (or Helmholtz) systems $(\Delta'_n+V'({\bf x}))\Psi=E\Psi$ can be incorporated into a single Laplace equation, and we can exploit the relationship between them. Second, via  a gauge transformation we can always transform the Laplace problem to flat space and make direct use of the conformal symmetry algebra $so(n+2,C)$ of the Laplacian. Using this approach, we will show that the problem of classifying all 3D Helmholtz superintegrable systems with nondegenerate potentials, i.e., potentials with a maximal number of independent paprameters,  can be reduced to the problem of classifying the orbits of the nonlinear action of the conformal group on a 10-dimensional manifold.  Eventually, this should lead to a new classification structure for special functions and their properties.

\section{The classical Laplace-type conformal superintegrable system}
We start by defining Laplace-type conformal superintegrability in classical mechanics. The Hamiltonian system is  ${\cal H}=0$ where $ {\cal H}={\cal H}_0+V({\bf x})$ and ${\cal H}_0=\sum_{i,j=1}^ng^{ij}({\bf x})p_ip_j$ is the free particle Hamiltonian on a real or complex conformally flat pseudo-Riemannian space. The phase space is $2n$ dimensional with local coordinates $$({\bf x}, {\bf p})=(x_1,\cdots,x_n, p_1,\cdots,p_n).$$ The Poisson bracket  of functions $f,g$ on the phase space is
$$\{f,g\}=\sum_{i=1}^n(\partial_{x_i}f\partial_{p_i}g-\partial_{p_i}f\partial_{x_i}g)
$$
The condition ${\cal  H }=0$ restricts us to a $2n-1$ dimensional hypersurface in phase space.
A conformal symmetry of this equation is a function ${\cal S}({\bf x}, {\bf p})$ such that $\{ {\cal S},{\cal H}\}=R_{\cal S}({\bf x}, {\bf p} ){\cal H}$ for some function $R_{\cal S}$. Two conformal symmetries ${\cal S}, {\cal S}'$ are identified if ${\cal S}= {\cal S}'+R{\cal H}$ for $R$ any function on phase space, since they agree on the hypersurface ${\cal  H }=0$. The system is {\it conformally superintegrable} if there are $2n-1$ functionally independent conformal symmetries, ${\cal S}_1,\cdots,{\cal S}_{2n-1}$ with ${\cal S}_1={\cal H}$  which firstly, are polynomial in the momenta and secondly, the symmetries ${\cal S}_2,\cdots,{\cal S}_{2n-1}$ are still functionally independent on restriction to the hypersurface ${\cal H}=0$. The system is second order conformally superintegrable if each of the basis symmetries ${\cal S}_i$ can be chosen as a second order polynomial in the momenta.
The condition that $\cal S$ is a conformal symmetry implies $d {\cal S}/dt=\frac12\{{\cal S},{\cal H}\}=\frac12R_{\cal S}{\cal H}$  so $d {\cal S}/dt=0$ at  any point $({\bf x},{\bf p})$ on the hypersurface ${\cal H}=0$ and {\cal  S} is constant along any trajectory satisfying Hamilton's equations 
$$2\partial_t {\bf x} =\partial_{\bf p} {\cal H},\  2\partial_t{\bf  p} =-\partial_{\bf x} {\cal H}.$$
Note that if a point of the trajectory lies on the hypersurface ${\cal H}=0$ then all points on the trajectory lie on this hypersurface. Thus for constants $ c=(c_i )$, with $c_1=0$ we can solve the equations ${\cal S}_i({\bf x},\ {\bf p})=c_i$, $i=1,\cdots,2n-1$  analytically to get a 1-parameter  trajectory.

\section{The operator Laplace-type conformal superintegrable system} 
Systems of  Laplace type are of the form 
 \be\label{Laplace} H\Psi\equiv \Delta_n\Psi+V\Psi=0.\ee
 Here $\Delta_n $ is the Laplace-Beltrami operator on a real or complex conformally flat Riemannian or pseudo Riemannian manifold.
A conformal symmetry of this equation is a partial differential operator  $ S$ in the variables ${\bf x}=(x_1,\cdots,x_n)$  such that $[ S, H]\equiv SH-HS=R_{ S} H$ for some differential operator  $R_{S}$. A conformal symmetry maps any solution $\Psi$ of (\ref{Laplace}) to another solution. Two conformal symmetries ${ S}, { S}'$ are identified if $S=S'+RH$ for some differential operator $R$, since they agree on the solution space of (\ref{Laplace}). The system is {\it conformally superintegrable} if there are $2n-1$ functionally independent conformal symmetries, ${ S}_1,\cdots,{ S}_{2n-1}$ with ${ S}_1={ H}$. It is second order conformally superintegrable if each  symmetry $S_i$ can be chosen to be a  differential operator of at most second order.

We can distinguish three types of conformally superintegrable Laplace equations. The first type is just a recasting of a  Helmholtz superintegrable system $H'\Psi=E\Psi$ into Laplace form $H\Psi=0$ where  $H=H'-E$. Thus the parameter $E$ is absorbed into the potential. The second type is a restriction of a Helmholtz superintegrable system $H'\Psi=E\Psi$ where 
$H' = \Delta_n+V$ to a fixed energy eigenvalue $E_0$ The resulting system 
$$ \left(\Delta_n+V({\bf x})-E_0\right)\Psi=0
$$ 
is trivially conformally superintegrable. Here $E_0$ is a fixed constant in the potential ${\tilde V}=V-E_0$, not a parameter, and it is not permitted to add further nonzero constants to the potential,  \cite{CR}. Generically, all of the symmetries of a system of this type will be those inherited from the Helmholtz equation, so the restriction seems to be no special interest. However, for some particular energies $E_0$ new truly conformal symmetries may appear so that the structure of the symmetry algebra will change.
The third type of conformal superintegrable system is one of the form (\ref{Laplace}) where the Helmholtz equation $H\Psi=E\Psi$
is not superintegrable. In this case the truly conformal symmetries necessarily appear.

\subsection{A Laplace equation example with degenerate potential}
For our first example we consider the equation
 \be \label{Laplace1}H\Psi\equiv\sum_{i=1}^n(\partial_{x_i}^2+\frac{a_i}{x_i^2})\Psi=0.\ee
(This equation was  treated by Volkmer  in his study of generalized ellipsoidal harmonics in n dimensional Euclidean space, \cite{Volkmer2006}.)  In this case the Helmholtz equation $H\Psi=E\Psi$ is second order superintegrable with degenerate potential. A basis of generators for the quadratic algebra of symmetries is given by the $n(n+1)/2$ second order symmetries
\be\label{Pj'} P_j=\partial^2_{x_j}+\frac{a_j}{x_j^2},\quad j=1,\cdots,n\ee
and 
\be \label{Jjk'} J_{jk}=(x_j\partial_{x_k}-x_k\partial_{x_j})^2+a_j\frac{x_k^2}{x_j^2}+a_k\frac{x_j^2}{x_k^2},\quad 1\le j<k\le n.\ee
(Of course there are functional relations between these symmetries since only a $2n-1$ element subset is functionally independent. For the case $n=3$ these relations can be found in \cite{KMPost4}, by restriction.) What makes this potential of particular interest is that for $E=0$ the system admits new (truly conformal) symmetries. The most obvious is the dilation symmetry
\be\label{dilation} D=-\sum_{i=1}^n x_i\partial_{x_i}-\frac{n-2}{2}.\ee 
(The constant term has been added for convenience in the calculations to follow.)  Note that $[D,H]=2H$. Further, there is a non-local symmetry $I$ defined by
\be\label{Isymmetry} I\Psi (x,y,z)=\frac{1}{r}\Psi\left(\frac{x}{r^2},\frac{y}{r^2},\frac{z}{r^2}\right),\quad r^2=\sum_{i=1}^nx^2_i.\ee
Here,
$$ [I,H]=r^4 H,\quad I=I^{-1},$$
so if $S$ is a differential symmetry, so is $ISI^{-1}$. Now $IJ_{jk}I^{-1}=J_{jk}$ and $IDI^{-1}=-D$, so we get nothing new. However the operators $K_j=IP_jI^{-1}$ are new conformal symmetries:
\be\label{K_j'} K_j=\left((n-2)x_j-\sum_{i=1}^nx^2_i\partial_{x_j}+2x_j\sum_{i=1}^n x_i\partial_{x_i}\right)^2+a_j\frac{(\sum_{i=1}^n x_i^2)^2}{x_j^2}.\ee
These symmetries are not independent of one another. In particular we have the identities
\be\label{conformalident1} \sum_{i=1}^n P_i=0,\ \sum_{i=1}^n K_i=0,\ \sum_{1\le j<k\le n}J_{jk}+D^2+\sum_{i=1}^n a_i-\frac{n-2}{2}=0,
\ee
each valid on the solution space of $H\Psi=0$. The first order conformal symmetry $D$ acts on the second order symmetries via
\be\label{Daction} [D,P_j]=2P_j,\ [D,K_j]=-2K_j,\ [D,J_{jk}]=0.
\ee
We also have the second order commutator relations
$$[P_i,P_j]=0,\ [K_i,K_j]=0.$$
The expressions for the commutators $[P_i,J_{jk}],\ [K_i,J_{jk}]$ and $[P_i,K_j]$ are more complicated and the structure of the symmetry algebra generated via commutation is not completely clear at this time. Note however, since (\ref{Laplace}) can be thought of as a restriction of the singular isotropic oscillator for the Helmholtz equation in $n$ dimensions, the operators $[P_i,J_{jk}][P_{i'},J_{j'k'}] +[P_{i'},J_{j'k'}][P_i,J_{jk}]$ can be expressed as symmetrized third order polynomials in $P_\ell$ and $J_{hm}$. Similarly by applying the symmetry $I$ we see that  the operators $[K_i,J_{jk}][K_{i'},J_{j'k'}] +[K_{i'},J_{j'k'}][K_i,J_{jk}]$ can be expressed as symmetrized third order polynomials in $K_\ell$ and $J_{hm}$.  Using the same reasoning we see that fourth order operators of the form $[[P_i,J_{jk}],J_{j'k'}]$ or $[[P_i,J_{jk}],P_{i'}]$ can be expressed as symmetrized second order polynomials in $J_{j'',k''}$ and $P_{i''}$. Similarly fourth order operators of the form $[[K_i,J_{jk}],J_{j'k'}]$ or $[[K_i,J_{jk}],K_{i'}]$ can be expressed as symmetrized second order polynomials in $J_{j'',k''}$ and $K_{i''}$.

It is important to recognize that for eigenfunctions $\Psi_\lambda$ of the dilation operator, $D\Psi_\lambda=\lambda\Psi_\lambda$, the conformally superintegrable system on flat space specializes to the Helmholtz equation on the $n-1$ sphere  with generic potential:
\be\label{Helmholtzsphere}\sum_{1\le j<k\le n}J_{jk}\Psi_\lambda=(-\lambda^2+\frac{n-2}{2})\Psi_\lambda,\ee
a superintegrable system. Thus the  conformal symmetry algebra of (\ref{Laplace1}) can be regarded as a dynamical symmetry algebra for (\ref{Helmholtzsphere}) since in general these conformal symmetries will change the eigenvalue $\lambda$, hence the energy. Since the operators $[P_i,K_j]$ commute with $D$, they are symmetries of (\ref{Helmholtzsphere}) hence expressible in terms of the commutators of the basis symmetries $J_{jk}$ for the $n-1$ sphere with generic potential. Similarly, since symmetries of the form $[P_i,J_{jk}][K_{i'},J_{j'k'}]+[K_{i'},J_{j'k'}][P_i,J_{jk}]$ also commute with $D$ they must be expressible as third order symmetrized polynomials in the basis symmetries $J_{jk}$. In like manner, fourth order symmetries for the form $[ [P_i,J_{jk}],K_\ell ]$ and any like expressions that commute with $D$ must also be expressible as symmetrized second order polynomials in the $J_{j'k'}$.

Putting these observations together, we conclude that the conformal symmetry algebra of (\ref{Laplace1}) generated by $ P_i,J_{jk},K_\ell,D $ must close at order six, so it is a true quadratic conformal symmetry algebra. The structure theory for the quadratic conformal symmetry algebra and the complete set of functional relations among the generators is yet to be determined. However, for $n=2,3$ we can understand the functional relations. 

For $n=2$, the simplest and atypical case, there are 3 functionally independent generators, whereas we have 6 second order conformal symmetries $P_1,P_2,J_{12},K_1,K_2,D^2$. The relations
$$ P_1+P_2= H\sim 0,\quad K_1+K_2=(x_1^2+x_2^2)^2H\sim 0,$$
$$ J_{12}+D^2+a_1+a_2=(x_1^2+x_2^2)H\sim 0,$$
 \be \label{n=2relations}J_{12}^2-\frac12(K_1P_1+P_1K_1)-5J_{12}-3(a_1 +a_2 )-4a_1 a_2= \ee
$$-
\left(x^2(x^2-2y^2)\partial_x^2-x^4\partial_y^2 
+4x^3y\partial_{xy} \nonumber\right.$$
 $$\left. + x(6x^2 -4y^2 )\partial_x+6x^2 y\partial_y+9x^2 
 +a_1x^2 y^2 +2a_1y^2 -a_2\frac{x^4}{y^2}\right)H\sim 0 $$
  yield the complete structure, where we write $A\sim B$ if the operators $A,B$ have the same action on the null space of $H$. Indeed, we can take $H,P_1,K_2$ as the functionally independent generators. Then we have commutation relations  (\ref{Daction}) and
$$ [P_1,K_1]\sim D^3+4(1+2a_1+2a_2)D$$
which determine everything. In this case the algebra closes at order 3.

 If we further set $a_2=0$ then the structure of the conformal symmetry algebra changes again. We get a new first order symmetry $L_1=\partial_{x_2}$ and, since $K_2=L^2_2$ becomes a perfect square, another first order conformal symmetry 
$$L_2=-(x^2_1+x^2_2)\partial_{x_2}+2x_2(x_1\partial_{x_1}+x_2\partial_{x_2}).$$
Thus the system is now first order superintegrable with structure
$$ [D,L_1]=L_1,\ [D,L_2]=-L_2,\ [L_1,L_2]=-2D,$$
the Lie algebra $s\ell(2)$. To within a gauge transformation the Laplace equation $H\Psi=0$ is just a complexification of the EPD equation, studied in  \cite{KM76} from the group theoretic point of view.

For $n=3$ there are 10 second order symmetries $P_1,P_2,P_3,K_1,K_2,K_3$, $J_{12},J_{13},J_{23},D^2$, only 5 of which are functionally independent. This is explained by the 3 second order relations (\ref{conformalident1}) and 2 eighth order relations, one relating the $P_i,J_{jk}$ and one relating the $K_i,J_{jk}$. The algebra closes at order 6, as it does for all $n>2$.

\subsection{ Laplace equation examples with nondegenerate potential}
\begin{example}
We choose three dimensional  flat space with Cartesian coordinates $x,y,z$, though the construction works in all higher dimensions. We set
$$x=\frac{s_1}{1+s_4},\ y=\frac{s_2}{1+s_4},\  z=\frac{s_3}{1+s_4},\quad s_1^2+s_2^2+s_3^2+s_4^2=1,$$
essentially the stereographic projection of the three-sphere on Euclidian space.
We now choose generic coordinates on the three-sphere that correspond to $R$-separable coordinates for the flat space Laplace equation, as in \cite{Miller77}:
$$ s_1=\sqrt{\frac{(\mu-a)(\nu-a)(\rho-a)}{(b-a)(a-1)a}},\ s_2=\sqrt{\frac{(\mu-b)(\nu-b)(\rho-b)}{(a-b)(b-1)b}},\  $$
\be \label{ellipsoidalcoords} s_3=\sqrt{-\frac{(\mu-1)(\nu-1)(\rho-1)}{(a-1)(b-1)}},\ s_4=\sqrt{-\frac{\mu\nu\rho}{ab}}.\ee
The metric is
\be\label{metric}ds^2=dx^2+dy^2+dz^2=\frac{1}{(1+s_4)^2}\left(ds_1^2+ds_2^2+ds_3^2+ds_4^2\right).\ee
We can invert the coordinate transformation according to
$$s_1=\frac{2x}{x^2+y^2+z^2+1},\ s_2=\frac{2y}{x^2+y^2+z^2+1},\ s_3=\frac{2z}{x^2+y^2+z^2+1},$$
$$ s_4=\frac{1-x^2-y^2-z^2}{x^2+y^2+z^2+1}.$$

If we proceed at the classical level, then by choosing the coordinates indicated above we
can achieve multiseparation and hence conformal superintegrability with  a potential $ V $ of
the form
\bea\label{nondegenpotential} V&=&(1+s_4)^2\left(\frac{a_1}{s_1^2}+\frac{a_2}{s_2^2}+\frac{a_3}{s_3^2}+\frac{a_4}{s_4^2}-a_5\right)\\
&=&\frac{a_1}{x^2}+\frac{a_2}{y^2}+\frac{a_3}{z^2}+\frac{4a_4}{(1-x^2-y^2-z^2)^2}-\frac{4a_5}{(1+x^2+y^2+z^2)^2}.\nonumber\eea
This is  a conformally superintegrable system with nondegenerate potential. For the quantum analogue we obtain
a quite specific spectrum of the corresponding Laplace operator on the three sphere
with eigenvalue shifted by $3/4$. Indeed, writing Laplace's equation 
\be\label{Laplace2}(\partial_{xx}+\partial_{yy}+\partial_{zz}+V)\Psi=0\ee
in terms of the ellipsoidal coordinates $\mu,\nu,\rho$ and applying the gauge transformation $\Psi=(1+s_4)^{-1/2}\Phi$ we find
that (\ref{Laplace2}) transforms to a separable system with
 separation equations 
$$(\lambda-a)(\lambda-b)(\lambda-1)\lambda\left(\partial^2_\lambda+\frac12(\frac{1}{\lambda-a}+\frac{1}{\lambda-b}+\frac{1}{\lambda-1}+\frac{1}{\lambda})\partial_\lambda-a_1\frac{(a-b)(a-1)a}{\lambda-a}\right.$$
$$\left.-a_2\frac{(b-a)(b-1)b}{\lambda-b}-a_3\frac{(1-b)(1-a)}{\lambda-1}+\frac{a_4ab}{\lambda}-(a_5+\frac34)\lambda^2+\kappa_1\lambda+\kappa_2\right)\Lambda(\lambda)=0$$
for $\lambda=\mu,\nu,\rho$ and separation constants $\kappa_1,\kappa_2$. In terms of the function $\Phi$ the Laplace equation becomes
\be\label{Laplace3} \left(\Delta_{S_3}+\frac{a_1}{s_1^2}+\frac{a_2}{s_2^2}+\frac{a_3}{s_3^2}+\frac{a_4}{s_4^2}-(a_5+\frac34)\right)\Phi=0\ee
where $\Delta_{S_3}$ is the Laplace-Beltrami operator on the three sphere. This clearly illustrates the connection with the superintegrable system with nondegenerate potential on the three sphere for the energy translated by the  specific value $3/4$. Here the $3/4$ is due to the quantization procedure. In $2$ dimensions it it doesn't appear and for $n>2$ dimensions it is $3/16$ of the scalar curvature of the manifold. The quadratic algebra structure for the system (\ref{Laplace3}) is preserved by the gauge transformation to give the conformal quadratic algebra for system (\ref{Laplace2}). This map from a nondegenerate Helmholtz superintegrable system on the 3-sphere with potential
$$ {\tilde V}=\frac{a_1}{s_1^2}+\frac{a_2}{s_2^2}+\frac{a_3}{s_3^2}+\frac{a_4}{s_4^2}-a_5$$
to a nondegenerate Laplace conformally superintegrable system in flat space with potential
$$ V= (1+s_4)^2{\tilde V}=(1+s_4)^2\left(\frac{a_1}{s_1^2}+\frac{a_2}{s_2^2}+\frac{a_3}{s_3^2}+\frac{a_4}{s_4^2}-a_5\right)$$
  is {\it not} a St\"ackel transform, \cite{BKM,HGDR}. (It is an example of a St\"ackel multiplier, \cite{KM81}.)  However, since $(1+s_4)^2$ is a specialization of the Laplace potential, the inverse  transform {\it can} be considered a St\"ackel transform. Further, we can apply  St\"ackel transforms to this example, and to our previous example, to obtain  multiple conformal superintegrable systems on manifolds with nonconstant curvature. This idea will be used to characterize arbitrary Helmholtz superintegrable systems on conformally flat spaces as St\"ackel transforms of flat space Laplace superintegrable systems.
\end{example}
\begin{example}
A second approach to example (\ref{nondegenpotential}) uses  general cyclidic coordinates $\rho,\mu,\nu$.  We 
choose 
$$x^2_1 = \frac{(\rho -e_1)(\mu -e_1)(\nu -e_1)}{ (e_1-e_2)(e_1-e_3)(e_1-e_4)(e_1-e_5)},\quad 
x^2_2 = \frac{(\rho -e_2)(\mu -e_2)(\nu -e_2)}{ (e_2-e_1)(e_2-e_3)(e_2-e_4)(e_2-e_5)},$$
$$x^2_3 = \frac{(\rho -e_3)(\mu -e_3)(\nu -e_3)}{ (e_3-e_2)(e_3-e_1)(e_3-e_4)(e_3-e_5)},\quad
x^2_4 = \frac{(\rho -e_4)(\mu -e_4)(\nu -e_4)}{ (e_4-e_2)(e_4-e_3)(e_4-e_1)(e_4-e_5)},$$
$$x^2_5 = \frac{(\rho -e_5)(\mu -e_5)(\nu -e_5)}{ (e_5-e_2)(e_5-e_3)(e_5-e_4)(e_5-e_1)}.$$
Here $e_1,\cdots,e_5$ are constants. These are the pentaspherical coordinates on the cone  
\be\label{nullcone}x^2_1+x^2_2+x^2_3+x^2_4+x^2_5=0.\ee
 and they can be written in terms of 
projective coordinates $X,Y,Z,T$
\be \label{projectivecords}x_1=2XT ,\quad x_2=2YT ,\quad x_3=2ZT ,\quad x_4=X^2+Y^2+Z^2-T^2 ,\ee
$$\quad x_5=i(X^2+Y^2+Z^2+T^2).$$
The Cartesian coordinates $x,y,z$ are given by 
\be\label{pentasphericalcoords}x=\frac{X}{T}=-\frac{x_1}{x_4+ix_5},\quad y=\frac{Y}{T}=-\frac{x_2}{x_4+ix_5},\quad z=\frac{Z}{T}=-\frac{x_3}{x_4+ix_5}.\ee
We also note the relations  
\be \label{pentaspericalident} x^2+y^2+z^2-1=-\frac{2x_4}{x_4+ix_5},\quad x^2+y^2+z^2+1=\frac{2ix_5}{x_4+ix_5}.\ee

The metric distance in Euclidean space is 
\be\label{cycydicmetric}ds^2=dx^2+dy^2+dz^2=(x_4+ix_5)^{-2}\left[\frac{(\rho -\mu )(\rho -\nu )d\rho ^2}{ (\rho -e_1)(\rho -e_2)(\rho -e_3)(\rho -
e_4)(\rho -e_5)}+\right.\ee
$$ \left.\frac
{(\mu -\nu )(\mu -\rho )d\mu ^2}{ (\mu -e_1)(\mu -e_2)(\mu -e_3)(\mu -e_4)(
\mu -e_5)}+ \frac
{(\nu -\mu )(\nu -\rho )d\nu ^2}{ (\nu -e_1)(\nu -e_2)(\nu -e_3)(\nu -e_4)(
\nu -e_5)}\right].$$
We can now construct a general potential which is conformally superintegrable 
viz 
$$V=(x_4+ix_5)^2(\frac{a_1}{ x^2_1} +\frac {a_2}{ x^2_2} + \frac{a_3}{ x^2_3} + 
\frac{a_4}{ x^2_4} + \frac{a_5}{ x^2_5} )$$
which is identical to (\ref{nondegenpotential}) when written in terms of Cartesian coordinates.

Writing Laplace's equation  (\ref{Laplace2})
in terms of these coordinates and setting $\Psi =(x_4+ix_5)^{-1/2}\Phi $ we 
obtain a  partial differential equation for $\Phi$ that is separable: 
$\Phi(\rho,\mu,\nu)=\Lambda_1(\rho)\Lambda_2(\mu)\Lambda_3(\nu)$.
This leads to the three separation equations 
$$\left[4(\lambda -e_1)(\lambda -e_2)(\lambda -e_3)(\lambda -e_4)(\lambda -e_5)[
\partial ^2_\lambda +\right.$$
$$\left.
\frac12(\frac{1}{ \lambda -e_1}+ \frac{1}{\lambda -e_2}+ 
\frac{1}{ \lambda -e_3}+ \frac{1}{ \lambda -e_4}+ 
\frac{1}{ \lambda -e_5})\partial _\lambda ]\right.$$
$$+a_1\frac{(e_1-e_2)(e_1-e_3)(e_1-e_4)(e_1-e_5)}{ (\lambda -e_1)} + 
a_2\frac{(e_2-e_1)(e_2-e_3)(e_2-e_4)(e_2-e_5)}{ (\lambda -e_2)}$$
$$
+a_3\frac{(e_3-e_2)(e_3-e_1)(e_3-e_4)(e_3-e_5)}{ (\lambda -e_3)} + 
 a_4\frac{(e_4-e_2)(e_4-e_3)(e_4-e_1)(e_4-e_5)}{ (\lambda -e_4)}$$
$$\left.
+a_5\frac{(e_5-e_2)(e_5-e_3)(e_5-e_4)(e_5-e_1)}{ (\lambda -e_5)} -\frac54
\lambda ^3+\right.$$
$$\left.\frac34(e_1+e_2+e_3+e_4+e_5)\lambda ^2+\kappa _1\lambda +\kappa _
2\right]\Lambda (\lambda )=0,$$
$\lambda=\rho,\mu,\nu$.

If we  take $e_5\rightarrow  \infty $ then we recover the system above 
associated with the three dimensional sphere. In particular we note that the 
expression for the gauge factor is readily computed. This is a consequence of 
the relation  
$e_1x^2_1+e_2x^2_2+e_3x^2_3+e_4x^2_4+e_5x^2_5=-1$,
from which we deduce that  
$$-(x_4+ix_5)^2 
=e_1x^2+e_2y^2+e_3z^2+ e_4(1-x^2-y^2-z^2)^2-e_5(1+x^2+y^2+z^
2)^2.$$
\end{example}
\begin{example}
We consider the nondegenerate Laplace conformally superintegrable system in flat space with potential
\be\label{nondegenpotential1}
V(x,y,z)=\frac{a_1}{(x+iy)^2}+\frac{a_2z}{(x+iy)^3}+\frac{a_3(x^2+y^2-3z^2)}{(x+iy)^4}+\frac{a_4}{(1-x^2-y^2-z^2)^2}\ee
$$+\frac{a_5}{(1+x^2+y^2+z^2)^2}.$$
(This system will be real in real Minkowski space with coordinates $X=x$, $Y=iy$, $Z=z$.) A specialization of this potential is $1/(1+x^2+y^2+z^2)^2$ and if we use it to perform a St\"ackel transform we get a Helmholtz superintegrable system on the complex 3-sphere, in complete analogy with our first example. This is the system IV' in \cite{KKM20052}. Another specialization of potential 
(\ref{nondegenpotential1}) is $1/(x+iy)^2$. If we use it to perform a St\"ackel transform we get a Helmholtz superintegrable system on complex flat space,  the system IV in \cite{KKM20052}.

Example (\ref{nondegenpotential1}) is just one of a class of nondegenerate supperintegrable systems on the 3-sphere and flat space that are induced from nondegenerate superintegrable systems on the 2-sphere. Indeed, any superintegrable system on the 2-sphere can be imbedded in 3 dimensional flat space in an obvious manner. For each  3-parameter potential of a 2-sphere nondegenerate superintegrable system, such as listed in \cite{KKMP2001}, we get a flat space superintegrable system whose coefficients of $a_1,a_2,a_3$ can be read off from the list in \cite{KKMP2001}. Then we add the terms $a_4/(1-x^2-y^2-z^2)^2+a_5/(1+x^2+y^2+z^2)^2$ to get the corresponding Laplace nondegenerate superintegrable system.
\end{example}
\subsection{Pentaspherical coordinates and linearization of the conformal group}
Here we examine the role of pentaspherical coordinates (\ref{nullcone}), (\ref{projectivecords}), (\ref{pentasphericalcoords}), (\ref{pentaspericalident}) in more detail. 
From these relations we can write  
$$\partial _X=2T\partial _{x_1}+2X\partial _{x_4}+2iX\partial _{x_5},\ 
\partial _Y=2T\partial _{x_1}+2Y\partial _{x_4}+2iY\partial _{x_5},$$
$$\partial _Z=2T\partial _{x_1}+2Z\partial _{x_4}+2iZ\partial _{x_5}.$$
We also recognize 
$\partial _x=T\partial _X$ , $\partial _y=T\partial _Y$ , $\partial _z=T\partial _Z$.
From these observations we deduce that  the spatial derivatives are related to the pentaspherical derivatives via
$$\partial _x=-(x_4+ix_5)\partial _{x_1}+x_1(\partial _{x_4}+i\partial _{x_5}),\
\partial _y=-(x_4+ix_5)\partial _{x_2}+x_2(\partial _{x_4}+i\partial _{x_5}),\ $$
$$\partial _z=-(x_4+ix_5)\partial _{x_3}+x_3(\partial _{x_4}+i\partial _{x_5}).$$
The classical analogs of these relations are 
$$p_x=-(x_4+ix_5)p_{x_1}+x_1(p_{x_4}+ip_{x_5}),\
p_y=-(x_4+ix_5)p_{x_2}+x_2(p_{x_4}+ip_{x_5}),\ $$
$$
p_z=-(x_4+ix_5)p_{x_3}+x_3(p_{x_4}+ip_{x_5}).$$

From these relations we determine that the flat space free Hamiltonian is 
\be\label{pentfreeHam}{\cal H}_0=p^2_x+p^2_y+p^2_z = 
(x_4+ix_5)^2(p^2_{x_1}+p^2_{x_2}+p^2_{x_3}+p^2_{x_4}+p^2_{x_5}).\ee
Recalling that motion is restricted to the null cone
$x^2_1+x^2_2+x^2_3+x^2_4+x^2_5=0$, we can consider ${\cal H}_0$ as a Hamiltonian in 10-dimensional pentaspherical phase space with  Poisson bracket $\{\cdot,\cdot\}_P$ with respect to which  $x_i, p_{x_i}$ are canonically conjugate variables only if 
$$\{\sum_{k=1}^5x_k^2,{\cal H}_0\}_P=2(x_4+x_5)^2\sum_{k=1}^5x_kp_{x_k}=0,$$
so we restrict our consideration to the subspace of pentaspherical phase space that is on the null cone and for which
\be\label{pentrestrict} x_1p_{x_1}+x_2p_{x_2}+x_3p_{x_3}+x_4p_{x_4}+x_5p_{x_5}=0.\ee
The point of all this is that the action of conformal   symmetries is linearized in pentaspherical phase space. Indeed, at the conformal Killing vector level we have
$$ p_{x}= (x_1p_{x_4}-x_4p_{x_1})+i(x_1p_{x_5}-x_5p_{x_1}),\  p_{y}= (x_2p_{x_4}-x_4p_{x_2})+i(x_2p_{x_5}-x_5p_{x_2}), $$
$$ p_{z}= (x_3p_{x_4}-x_4p_{x_3})+i(x_3p_{x_5}-x_5p_{x_3}),$$
$$xp_y-yp_x=x_1p_{x_2}-x_2p_{x_1},\ yp_z-zp_y=x_2p_{x_3}-x_3p_{x_2},\ $$
$$ zp_x-xp_z=x_3p_{x_1}-x_1p_{x_3},\ xp_x+yp_y+zp_z=i(x_4p_{x_5}-x_5p_{x_4}),$$
$$-(x^2+y^2+z^2)p_x+2x(xp_x+yp_y+zp_z)=(x_1p_{x_4}-x_4p_{x_1})-i(x_1p_{x_5}-x_5p_{x_1}),\ $$
$$
-(x^2+y^2+z^2)p_y+2y(xp_x+yp_y+zp_z)=(x_2p_{x_4}-x_4p_{x_2})-i(x_2p_{x_5}-x_5p_{x_2}),$$
$$-(x^2+y^2+z^2)p_z+2z(xp_x+yp_y+zp_z)=(x_3p_{x_4}-x_4p_{x_3})-i(x_3p_{x_5}-x_5p_{x_3}).$$
This means that the classical conformal Killing vectors  of ${\cal H}_0$ are just those of the complex Lie 
algebra of $SO(5,{\bf C})$, and the corresponding connected component to the identity of the  group symmetries is just  this linear Lie group.  Classically, the inversion operation in a sphere corresponds to the reflection 
\be\label{inversion} I:\ x_j\to x_j,\ p_{x_j}\to p_{x_j},\ j\ne 4,\quad x_4\to -x_4.\ p_{x_4}\to -p_{x_4}\ee
Thus the full conformal group action in pentaspherical space is just the linear action of $O(5,{\bf C})$. 

A straightforward computation shows that under the null cone restriction (\ref{nullcone}) and the restriction (\ref{pentrestrict}) the canonical one-form $\omega=p_xdx+p_ydy+p_zdz$ on our original phase space goes to the  canonical one-form on the restricted pentaspherical space:
$$\omega=p_xdx+p_ydy+p_zdz=\sum_{k=1}^5p_{x_k}dx_k.$$
Thus 
$$ d\omega = dp_x\wedge dx+dp_y\wedge dy+dp_z\wedge dz=\sum_{k=1}^5dp_{x_k}\wedge  dx_k$$
so the symplectic two-forms agree and we have achieved an embedding of our 6-dimensional phase space into the 10-dimensional pentaspherical phase space that preserves Poisson bracket relations.

The flat space classical system ${\cal H}=p_x^2+p_y^2+p_z^2+V({\bf x})=0$ can now be lifted to pentaspherical space:
$$
{\cal H}=(x_4+ix_5)^2(\sum_{k=1}^5p_{x_k}^2)+V({\bf x})=0$$
where now $V({\bf x})=(x_4+ix_5)^2{\tilde V}(x)$ is expressed in pentaspherical coordinates $x_k$. Thus we have the equation
$$ \sum_{k=1}^5p_{x_k}^2+{\tilde V}(x)=0,$$
and the possibilities for $\tilde V$ to correspond to a superintegrable system relate to properties of confocal quadratic forms in 5-space. This construction was exploited long ago by B\^ocher in his monograph on $R$-separation of variables for Laplace equations \cite{Bocher}.
The general cyclidic coordinates and the coordinates on 
the sphere given previously are easily related to this construction. To 
see this observe that general cyclidic coordinates in 5-space are given by
$$x^2_1= 
\frac{(\rho -e_1)(\nu -e_1)(\mu -e_1)}{ (e_1-e_2)(e_1-e_3)(e_1-e_4)(e_1-e_5)},$$
with analogous expressions for $x_2,\cdots,x_5$.
Now take $e_1=0$ and substitute $\lambda \rightarrow 1/\lambda  
,\lambda =\rho ,\mu ,\nu $ and $e_i\rightarrow 1/e_i ,\ i=2,3,4,5$. We obtain 
$$x^2_1= \frac{\mu \nu \rho }{ e_2e_3e_4e_5},\ 
x^2_2=x^2_1 
\frac{(\mu -e_2)(\nu -e_2)(\rho -e_2)}{ (e_2-e_3)(e_2-e_4)(e_2-e_5)},\ 
x^2_3=x^2_1 
\frac{(\mu -e_3)(\nu -e_3)(\rho -e_3)}{(e_3-e_2)(e_3-e_4)(e_3-e_5)},$$
$$x^2_4=x^2_1 
\frac{(\mu -e_4)(\nu -e_4)(\rho -e_4)}{ (e_4-e_2)(e_4-e_3)(e_4-e_5)},\
x^2_5=x^2_1 
\frac{(\mu -e_5)(\nu -e_5)(\rho -e_5)}{ (e_5-e_3)(e_5-e_4)(e_5-e_2)}.$$
From this calculation we could just as well choose new pentaspherical 
coordinates  
$$X^2_1=-1,\ 
X^2_2= - \frac{(\mu -e_2)(\nu -e_2)(\rho -e_2)}{ (e_2-e_3)(e_2-e_4)(e_2-e_5)},\
X^2_3= - \frac{(\mu -e_3)(\nu -e_3)(\rho -e_3)}{ (e_3-e_2)(e_3-e_4)(e_3-e_5)},$$
$$X^2_4= - \frac{(\mu -e_4)(\nu -e_4)(\rho -e_4)}{ (e_4-e_3)(e_4-e_2)(e_4-e_5)},\
X^2_5= - \frac{(\mu -e_5)(\nu -e_5)(\rho -e_5)}{ (e_5-e_3)(e_5-e_4)(e_5-e_2)},$$
from which it is clear that $X^2_2+X^2_3+X^2_4+X^2_5=1$. If we now relabel these
coordinates according to $X_4=y_3,\ X_5=y_4,\ X_3=y_2,\ X_2=y_1, \ X_1=y_5$ and choose 
Cartesian-like coordinates according to 
$$x= - \frac{y_1}{ y_4+iy_5},\ 
y= - \frac{y_2}{ y_4+iy_5},\ 
z= - \frac{y_3}{ y_4+iy_5},$$
we obtain the separable coordinate systems on the sphere displayed previously. What this demonstrates
is that the Cartesian-like coordinates are determined only by the ratios of the 
pentaspherical coordinates as indicated above. In addition the conformal 
symmetry group is available, acting on the vector $x=(x_1,x_2,x_3,x_4,x_5)$ in 
the usual linear way. Observe that  if we have a pentaspherical vector $x$ 
with  nonzero length that corresponds to the origin of the underlying cartesian-like coordinate system, a  $O(5,{\bf C})$ rotation can reduce it 
to the form $x^0=(0,0,0,0,a)$. If $x$ is  of zero length then we can 
reduce it to the form $x^0=(0,0,0,a,-ia)$ where  $a$ is 
arbitrary. Consequently if we take the first case and choose $a= -i$,  the
pentaspherical coordinates $(x_1,x_2,x_3,x_4,-i)$ map 1-1 to the underlying space via 
$$x=\frac {X}{ T} = - \frac{x_1}{ x_4+1},\  y = \frac{Y}{ T}= - \frac{x_2}{ x_4+1} ,\ z = 
\frac{Z}{ T}= - \frac{x_3}{ x_4+1}$$
where $x^2_1+x^2_2+x^2_3+x^2_4=1$,  i.e., the unit 3-sphere. For
the second case we obtain, choosing $a = \frac{1}{ 2}$ that the pentaspherical coordinates $(x_1,x_2,x_3,1/2,-i/2)$ map 1-1 to the underlying space via 
$$x=x_1 ,\ y=x_2 ,\ z=x_3$$
which is tantamount to choosing Euclidean Cartesian coordinates. To 
construct superintegrable systems we can invoke what we already know about 
Euclidean space and the three dimensional sphere. There remains the question of 
equivalence however.

 If we now look at the corresponding problem for the Laplace
equation there are some modifications necessary. Using the same correspondence 
for pentaspherical coordinates as above, it follows from the work of B\^ocher \cite{Bocher} that if $\Psi $ is a 
solution of the  Laplace equation 
$$(\partial ^2_x+\partial ^2_y+\partial ^2_z+V)\Psi =0$$
and
 if we write $\Psi =(x_4+ix_5)^{1/2}\Phi(x,y,z)$, $V=(x_4+ix_5)^2{\tilde V}(x)$ the function $\Phi$ satisfies  
$$\left(\partial ^2_{x_1}+\partial ^2_{x_2}+\partial ^2_{x_3}+\partial ^2_{x_4}+
\partial ^2_{x_5}+{\tilde V}(x)\right)\Phi=0$$
and  has degree of homogeneity $-{1}/{ 2}$,  i.e.,
$$(x_1\partial _{x_1}+x_2\partial _{x_2}+x_3\partial _{x_3}+x_4\partial _{x_4}+x
_5\partial _{x_5})\Phi=-\frac{1}{ 2} \Phi.$$
Similarly, the function  $\tilde V$ has degree of homgeneity $-2$.
Since
$\Psi(x,y,z)=\Psi(- \frac{x_1}{ x_4+ix_5}$ , $- \frac{x_2}{ x_4+ix_5}$ , $- 
\frac{x_3}{x_4+ix_5})$,
 the comments made previously about the vector $x^0$ can be 
applied. Indeed if we take the case of the three dimensional sphere, the Laplace operator  acting on 
the function $\Phi$ yields the equation
$(\Delta _3+{\tilde V}-\frac{3}{ 4})\Phi=0$.
Consequently any potential that enables superintegrability to occur on the 
complex three sphere also occurs for the corresponding Laplace 
equation. Similarly, it is clear that if we make the second choice for $x^0$ we 
return to superintegrable systems in complex flat space in three dimensions. It remains to be shown whether these are
 all the possibilities.

To address this, note that the  symmetry group acts on the the functions $\Phi$ via the normal algebra 
operators $I_{ij}=x_i\partial _{x_j}-x_j\partial _{x_i}$
and via the correct multiplier representation on the functions $\Psi$. In the 
 book \cite{Bocher} there are mentioned possibilities that there are coordinate systems
corresponding to $S_2\times E_1$, (the two-sphere $\times$ the line). To see how these arise consider 
spheroidal coordinates  
$$x=c\sqrt {\rho \mu }\cos\varphi,\ y=c\sqrt {\rho \mu }\sin\varphi,\ z=c\sqrt {-(\rho -1)(\mu -1)}.$$
The corresponding infinitesimal distance is 
$$ds^2=c^2\left[\rho \mu d\varphi ^2-\frac14(\rho -\mu )[\frac{d\rho ^2}{ \rho (\rho -1)} - \frac{d\mu ^2}{ \mu (\mu -1)}]\right].$$
If we factor out $\rho \mu $ from this metric we see that 
$$ds^2=c^2\rho \mu ds'^2=c^2\rho \mu \left[d\varphi ^2-\frac{1}{ 4}(\rho -\mu )[\frac{d\rho 
^2}{\rho ^2(\rho -1)} - \frac{d\mu ^2}{ \mu ^2(\mu -1)}]\right]$$
where $ds'^2$ is clearly a metric corresponding to $S_2\times E_1$. From what we
have just done this is conformally equivalent to spheroidal coordinates. However it is also conformally equivalent to flat space. Similarly, all such $S_2\times E_1$ systems are conformally equivalent to systems on constant curvature spaces.

\subsection{Theory for 3D classical conformally 2nd order superintegrable Laplace systems}
Given a classical conformally superintegrable system on a conformally flat space we can always find a Cartesian like coordinate system with coordinates $(x,y,z)\equiv (x_1,x_2,x_3)$ such that the Hamilton-Jacobi (Laplace) equation takes the form
\be\label{Laplace4} \frac{p_1^1+p_2^2+p_3^2}{\lambda({\bf x})}+{\tilde V}({\bf x})=0.\ee
However, this equation is equivalent to the flat space equation
\be\label{Laplace5}{\cal H}\equiv  p_1^2+p_2^2+p_3^2+ V({\bf x})=0,\quad V({\bf x})=\lambda({\bf x}){\tilde V}({\bf x}).\ee
In particular, the conformal symmetries of (\ref{Laplace4}) are identical with the conformal symmetries of (\ref{Laplace5}). Thus without loss of generality in the classical case, we can assume the manifold is flat space with $\lambda\equiv 1$. (In the quantum case a similar result is true but a gauge transformation is required and the modification of the potential is dependent on curvature.)

The operation of inversion in a sphere plays a special role in the classical theory. If $u({\bf x})$ is a solution of the Hamilton-Jacobi equation 
$$p_1^2+p_2^2+p_3^2+{ V}({\bf x})=0,  \quad p_i=\partial_i u,$$
then 
\be\label{Isymmetry1} {\tilde u}({\bf x})=Iu({\bf x})=u(\frac{{\bf x}}{r^2}),\quad r^2=x^2+y^2+z^2\ee
satisfies 
$$ {\tilde p}_1^2+{\tilde p}_2^2+{\tilde p}_3^2+\frac{1}{r^4}V(\frac{\bf x}{r^2})=0,\quad {\tilde p}_i=\partial_i{\tilde u}.$$ 

A second order conformal symmetry \be
{\cal S}=\sum ^3_{k,j=1}a^{kj}({\bf x})p_kp_j+W({\bf x})\equiv
{\cal L}+W,\quad a^{jk}=a^{kj} \label{constofmotc}
\ee
must satisfy   

\be\label{confsym}
\{{\cal S},{\cal H}\}=b({\bf x},{\bf p}){\cal H},\quad b=b_1({\bf x})p_1+b_2({\bf x})p_2+b_3({\bf x})p_3,
\ee
for some functions $b_1,b_2,b_3$.
Equating coefficients of monomials in the $p$'s we see that the conditions are 
\bea\label{killingtensors}
a_i^{ii}&=&2a_j^{ij}+a_i^{jj}=2a_k^{ik}+a_i^{kk}=\frac12 b_i,\nonumber\\
 a^{ij}_k+a^{ki}_j+a^{jk}_i&=&0,\quad i,j,k\ {\rm pairwise \ distinct}\noindent\label{symalgc}
\eea
and
\be
W_k=\sum_{s=1}^3a^{sk}
V_s+a_k^{kk}V,\quad k=1,2,3
.\label{potc}
\ee
(Here a subscript $j$ on $a^{\ell m}$, $V$ or $W$ denotes differentiation with respect to $x_j$.)
The requirement that $\partial_{x_\ell} W_j=\partial_{x_j}
W_\ell,\ \ell\ne j$ leads from
(\ref{potc}) to the
second order (conformal) Bertrand-Darboux  partial differential equations for the potential.
\be\label{BertrandDarboux}
\sum_{s=1}^3\left[V_{sj} a^{s\ell}-V_{s\ell} a^{sj}+
V_s\left(
(a^{s\ell})_j-(
a^{sj})_\ell\right)\right]+a^{\ell \ell}_\ell V_j-a^{jj}_jV_\ell +(a_{j\ell}^{\ell\ell}-a_{j\ell}^{jj})V=0.
\ee

Equations (\ref{killingtensors}) are exactly the equations for a second order conformal Killing tensor $a^{ij}$. Thus, necessary and sufficient conditions that ${\cal S}={\cal L}+W$ be a conformal symmetry are that $\cal L$ is a conformal Killing tensor, $W$ is a solution of equations (\ref{potc}) and $V$ satisfies the conformal Bertrand-Darboux equations.

The
conformal Killing tensors for flat space are  well known, e.g., \cite{Eastwood, Miller77}.
The space of conformal Killing tensors is infinite dimensional.
It is spanned by products of the conformal Killing vectors
$$ p_1,\ p_2,\ p_3,\  x_3p_2-x_2p_3,\ x_1p_3-x_3p_1,\ x_2p_1-x_1p_2,\
x_1p_1+x_2p_2+x_3p_3,
$$
$$
(x_1^2-x_2^2-x_3^2)p_1+2x_1x_3p_3+2x_1x_2p_2,\ 
(x_2^2-x_1^2-x_3^2)p_2+2x_2x_3p_3+2x_2x_1p_1,\ 
$$
$$
(x_3^2-x_1^2-x_2^2)p_3+2x_3x_1p_1+2x_3x_2p_2,
$$
and terms $g({\bf x},{\bf p})(p_1^2+p_2^2+p_3^2)$ where $g$ is an
arbitrary function.   For a given conformal superintegrable system only a finite dimensional space of conformal tensors occurs.  This is for two reasons. First the conformal Bertrand-Darboux equations restrict the allowed Killing tensors. Second, on the hypersurface ${\cal H}=0$ in phase space all symmetries $g({\bf x}){\cal H}$ vanish, so any two symmetries  differing by $g({\bf x}){\cal H}$ can be identified.

It is sometimes useful to pass to new variables 
$ a^{11},\  a^{24},\ a^{34}$, $a^{12},\ a^{13},\ a^{23}$
for the conformal Killing tensor, where $a^{24}=a^{22}-a^{11}$, $a^{34}=a^{33}-a^{11}$. Then we see that 
$  a^{24},\ a^{34}$, $a^{12},\ a^{13},\ a^{23}$
must be polynomials of order $\le 4$. (Thus by adding  $-a^{11}{\cal H}$ to the second order symmetry we can achieve $a^{11}=0$ for the new tensor with the same action on the hypersurface ${\cal H}=0$, without changing the 5 other variables.)

  For second order conformal superintegrabilty in 3D there must
be five functionally independent conformal constants of the motion (including the Hamiltonian itself). Thus  the Hamilton-Jacobi 
equation admits four   additional constants of the 
motion: 
\be \label{3dconst}
{\cal S}_h=\sum_{j,k=1}^3a^{jk}_{(h)}p_kp_j+W_{(h)}={\cal L}_h+W_{(h)},\qquad h=1,\cdots,4.
\ee
We assume that the four functions ${ {\cal S}}_h$ together with ${\cal H}$ are functionally
independent in the six-dimensional phase space, i.e., that the
differentials $d{ {\cal S}}_h,\ d{\cal H}$ are linearly independent. (Here the possible $V$  will always be assumed to form a vector space and we require functional independence 
for each such $V$ and the associated $W^{(h)}$. This means that we
also require that the five quadratic forms ${\cal L}_h,\ {\cal H}_0$ are 
functionally independent.) We say that the functions are {\it weakly
  functionally independent} if $d{ {\cal S}}_h,\ d{\cal H}$ are
linearly independent for nonzero potentials, but not necessarily for
the zero potential. Indeed for now, we will also require that the generating basis is functionally linearly independent.

We can write the system of conformal Bertrand Darboux equations  in the matrix
form $Cv={\tilde v}^{(1)}V_1+{\tilde v}^{(2)}V_2+{\tilde v}^{(3)}V_3+{\tilde v}^{(0)}V$, or
$$
\left(\ba{ccccc}0&a^{12}&a^{11}-a^{22}&a^{31}&-a^{32}\\a^{13}&0&-a^{23}&a^{21}&a^{11}-a^{33}\\a^{32}&-a^{32}&-a^{13}&a^{22}-a^{33}&a^{12}\ea\right)
\left(\ba{c}V_{33}-V_{11}\\V_{22}-V_{11}\\V_{12}\\V_{32}\\V_{31}\ea\right)=
$$
\be\label{fundeqns12}
  \left(\ba{c}a^{12}_1- a^{11}_2+a^{22}_2\\  a^{31}_1- a^{11}_3+a^{33}_3
\\ a^{31}_2-a^{21}_3\ea\right)V_1+ \left(\ba{c} a^{22}_1-a^{21}_2-a^{11}_1\\ a^{32}_1- a^{12}_3
\\ a^{32}_2- a^{22}_3+a^{33}_3\ea\right)V_2\ee
$$+\left(\ba{c} a^{32}_1-a^{31}_2\\ a^{33}_1- a^{13}_3-a^{11}_1
\\ a^{33}_2-a^{23}_3-a^{22}_2\ea\right)V_3+\left(\ba{c}a^{22}_{21}-a^{11}_{12} \\ a^{33}_{31}-a^{11}_{13}\\ a^{33}_{32}-a^{22}_{23}\ea\right)V.
$$
\begin{corollary}\label{corollary1} Suppose the set $\{{\cal H},{
{\cal S}}_1,\cdots,{ {\cal S}}_4\}$ is functionally linearly independent.
Then for general $x$ the $4\times 5$ matrix  
$$
A=\left(\ba{ccccc}
a^{33}_{(1)}-a^{11}_{(1)},&a^{22}_{(1)}-a^{11}_{(1)},&a^{12}_{(1)},&a^{31}_{(1)},&a^{32}_{(1)}\\
a^{33}_{(2)}-a^{11}_{(2)},&a^{22}_{(2)}-a^{11}_{(2)},&a^{12}_{(2)},&a^{31}_{(2)},&a^{32}_{(2)}\\a^{33}_{(3)}-a^{11}_{(3)},&a^{22}_{(3)}-a^{11}_{(3)},&a^{12}_{(3)},&a^{31}_{(3)},&a^{32}_{(3)}\\
a^{33}_{(4)}-a^{11}_{(4)},&a^{22}_{(4)}-a^{11}_{(4)},&a^{12}_{(4)},&a^{31}_{(4)},&a^{32}_{(4)}\ea\right)
$$
has rank 4, where the functions $a^{ij}_{(h)}({\bf x})$ are given by (\ref{3dconst}).
\end{corollary}

There are four sets of equations (\ref{fundeqns12}), one for each of the functionally independent symmetries (in addition to the Hamiltonian). We can write them as a single matrix equation
$ Bv=b$ where $B$ is $12\times 5$, $b$ is $12\times 1$ and 
$$
v=\left(\ba{c}V_{33}-V_{11}\\V_{22}-V_{11}\\V_{12}\\V_{32}\\V_{31}\ea\right).
$$
\begin{lemma} \label{lemma11} If the set $\{{\cal H},{
{\cal S}}_1,\cdots,{ {\cal S}}_4\}$ is functionally linearly independent, the matrix $B$ has rank 5.
\end{lemma}
The proof is the same as for the corresponding Helmholtz result in \cite{KKM20051}, with corrections in the introduction to \cite{KKM20061}.

By choosing a rank 5
minor of $B$ we can solve for $v$ and obtain a solution of the form
\be\label{veqn1a}\ba{lllll}
V_{22}&=&V_{11}&+&A^{22}V_1+B^{22}V_2+C^{22}V_3+D^{22}V,\\
V_{33}&=&V_{11}&+&A^{33}V_1+B^{33}V_2+C^{33}V_3+D^{33}V,\\
V_{12}&=& &&A^{12}V_1+B^{12}V_2+C^{12}V_3+D^{12}V\\
V_{13}&=& &&A^{13}V_1+B^{13}V_2+C^{13}V_3+D^{13}V\\
V_{23}&=& &&A^{23}V_1+B^{23}V_2+C^{23}V_3+D^{23}V\ea
\ee
If the augmented matrix $(B,b)$ has rank $r'>r$ then there will be
$r'-r$ additional conditions involving only derivatives less than second order.
 Here the $A^{ij},B^{ij},C^{ij},D^{ij}$ are
functions of $\bf x$ that can be calculated explicitly. For convenience we take $A^{ij}\equiv A^{ji}$, $B^{ij}\equiv B^{ji}$, $C^{ij}\equiv C^{ji}$, $D^{ij}\equiv D^{ji}$. 
 
Suppose now that the superintegrable system is such that $r'=r$ so
that relations (\ref{veqn1a}) are equivalent to $Bv=b$. Further, suppose the integrability conditions for system  (\ref{veqn1a})  are satisfied identically. 
In this case we say that the potential is {\it nondegenerate}. Otherwise the potential is {\it degenerate}. If $V$ is nondegenerate then at any point ${\bf x}_0$, where the $A^{ij}, B^{ij}, C^{ij}, D^{ij}$ are 
defined and analytic, there is a unique solution $V({\bf x})$ with
arbitrarily prescribed values of $V({\bf x}_0), V_1({\bf x}_0), V_2({\bf x}_0),
V_3({\bf x}_0), V_{11}({\bf x}_0)$. The points ${\bf x}_0$ are called {\it regular}. 
The points of singularity for the $A^{ij},B^{ij},C^{ij}, D^{ij}$ form a
manifold of dimension $<3$. Degenerate potentials depend on fewer
parameters. (For example, we could have 
 $r'=r$ but the integrability conditions are not satisfied identically. Or a first order conformal symmetry might exist and this would imply a linear condition on the first derivatives of $V$ alone.) 
 
 Note that for a nondegenerate potential the solution space of (\ref{veqn1a}) is exactly 5-dimensional, i.e. the potential depends on 5 parameters. Degenerate potentials depend on $<$ 5 parameters.

\subsection{The conformal St\"ackel transform}
We quickly review the concept of the St\"ackel transform \cite{BKM} and extend it to conformally superintegrable systems.
Suppose we have a second order conformal  superintegrable system 
\be\label{confl} {\cal H}=\frac{p_1^2+p_2^2+p^2_3}{\lambda(x,y,z)}+V(x,y,z)=0,\quad  {\cal H}={\cal H}_0+V
\ee
and suppose $U(x,y,z) $ is a particular solution of equations (\ref{veqn1a}), nonzero in an open set.  
\begin{theorem}\label{stackelt}
The transformed (Helmholtz) system  \be\label{helms} {\tilde {\cal  H}}=E,\quad 
{\tilde {\cal  H}}=(p_1^2+p_2^2+p^2_3)/{{\tilde \lambda}}+{\tilde V}
\ee
 with   potential ${\tilde V}(x,y,z)$
is truly  superintegrable, where
$${\tilde \lambda}=\lambda U,\  {\tilde V}=\frac{V}{U},$$
\end{theorem}

\medskip\noindent
PROOF: Let ${\cal S}=\sum a^{ij}p_ip_j+W={\cal S}_0+W$  be a second order conformal symmetry of $\cal H$ and ${\cal S}_U=\sum a^{ij}p_ip_j+W_U={\cal S}_0+W_U$  be the special case  that is in conformal involution with 
$(p_1^2+p_2^2+p^2_3)/{ \lambda}+ U$. Then  $$\{ {\cal S},{\cal H}\}=\rho_{{\cal S}_0}{\cal H},\quad  \{{\cal S}_U,{\cal H}_0+U\}=\rho_{{\cal S}_0}({\cal H}_0+U),$$
and ${\tilde{\cal S} }={\cal S}-\frac{W_U}{U}{\cal H}$
is a corresponding true symmetry of $\tilde {\cal H}$. Indeed, 
$$\{{\tilde{\cal S}},{\tilde {\cal H}}\}=\{{\cal S},\frac{\cal H}{U}\}-\{W_U\frac{\cal H}{U},\frac{\cal H}{U}\}$$
$$=\rho_{{\cal S}_0}\frac{\cal H}{U}-\frac{\cal H}{U^2}\{{\cal S},U\}-\frac{\cal H}{U}\{W_U.\frac{\cal H}{U}\}=\rho_{{\cal S}_0}\frac{\cal H}{U}-\frac{\cal H}{U^2}\rho_{{\cal S}_0}U=0.$$
 This transformation of second order symmetries preserves linear and functional independence. Thus the transformed system is Helmholtz  superintegrable. Q.E.D.

There is a similar result for first order conformal symmetries ${\cal L}=\sum a_i({\bf x})p_i$.
\begin{corollary} Let $\cal L$ be a first order conformal symmetry  of the superintegrable system (\ref{confl}) and suppose $U({\bf x}) $ is a particular solution of equations (\ref{veqn1a}), nonzero in an open set. Then $\cal L$ is a true symmetry of the Helmholtz superintegrable system (\ref{helms}): $\{{\cal L},{\tilde {\cal H}}\}=0$.
 \end{corollary}

\medskip\noindent   PROOF: By assumption, $\{{\cal L},{\cal H}\}=\rho_{\cal L}({\bf x}){\cal H}=\rho_{\cal L}({\cal H}_0+V$ where $\rho_{\cal L}$ doesn't depend on the momenta $p_j$. Thus, $\{{\cal L}, {\cal H}_0\}=\rho_{\cal L} {\cal H}_0, \{{\cal L},V\}=\rho_{\cal L} V$, so also $\{{\cal L},U\}=\rho_{\cal L} U$ Then 
$$\{{\cal L},{\tilde{\cal H}}\}=\{{\cal L},\frac{\cal H}{U}\}=\frac{1}{U}\{{\cal L},{\cal H}\}-\frac{\cal H}{U^2}\{{\cal L},U\}=\rho_{\cal L}(\frac{\cal H}{U}-\frac{\cal H}{U^2}U)=0.$$
Q.E.D.

These results show that any second order conformal Laplace superintegrable system admitting a nonconstant potential $U$ can be St\"ackel transformed to a Helmholtz superintegrable system. This operation is invertible, although the inverse mapping is not a St\"ackel transform (it takes true symmetries to conformal symmetries). By choosing all possible special potentials $U$ associated with the  fixed Laplace system (\ref {confl}) we generate the equivalence class of all Helmholtz superintegrable systems (\ref{helms}) obtainable through this process. As is easy to check,  any two Helmholtz superintegrable systems lie in the same equivalence class if and only if they are St\"ackel equivalent in the standard sense. All Helmholtz superintegrable systems are related to conformal Laplace systems in this way, so the study of all Helmholtz superintegrability on conformally flat manifolds can be reduced to the study of all conformal Laplace superintegrable systems on flat space.

Clearly, the analous results are true in all dimensions $n\ge 2$. For $n=2$ they also extend to  operator Laplace equtions without change; for $n=3$ they also extend but, as discussed above, the potentials must be modified under the St\"ackel transform to take scalar curvature into account.

\begin{example} Consider the degenerate Laplace system (\ref{Laplace1}) in three variables $x,y,z$. If we choose $U=1/z^2$ and perform the corresponding St\"ackel transform we obtain a Helmholtz superintegrable system on the 3-sphere with two-parameter potential. This follows immediately from the fact that the metric $(dx^2+dy^2+dz^2)/z^2$ corresponds to a space with nonzero constant cuvature.
 \end{example}

\subsection{The integrability conditions for the potential}
To determine the integrability conditions we first  introduce the dependent variables
$W^{(0)}$, $W^{(1)}=V_1$, $W^{(2)}=V_2$, 
$W^{(3)}=V_3$, 
$W^{(4)}=V_{11}$,  the vector
\be\label{wvector1a}{\bf w}^{\rm tr}=( W^{(0)},W^{(1)},W^{(2)}, W^{(3)},
W^{(4)}),
\ee
and the matrices
\be
{\bf A}^{(1)}=\left(\ba{rrrrr} 0&1&0&0&0\\ 0&0&0&0&1\\ D^{12}&A^{12}&B^{12}&C^{12}&0\\ D^{13}&A^{13}&B^{13}&C^{13}&0\\ D^{14}&A^{14}&B^{14}&C^{14}&B^{12}-A^{22}\ea\right),
\ee
\be
{\bf A}^{(2)}=\left(\ba{rrrrr} 0&0&1&0&0\\ D^{12}&A^{12}&B^{12}&C^{12}&0 \\ D^{22}&A^{22}&B^{22}&C^{22}&1 \\ D^{23}&A^{23}&B^{23}&C^{23}&0 \\ D^{24}&A^{24}&B^{24}&C^{24}&A^{12}\ea\right),
\ee
\be
{\bf A}^{(3)}=\left(\ba{rrrrr} 0&0&0&1&0\\ D^{13}& A^{13}&B^{13}&C^{13}&0 \\ D^{23}&A^{23}&B^{23}&C^{23}&0 \\ D^{33}&A^{33}&B^{33}&C^{33}&1 \\ D^{34}&A^{34}&B^{34}&C^{34}&A^{13}\ea\right),
\ee
where 
\bea
 A^{14}&=&A^{12}_2-A^{22}_1+B^{12}A^{22}+A^{12}A^{12}-B^{22}A^{12}-C^{22}A^{13}+C^{12}A^{23}-D^{22},\nonumber\\
B^{14}&=&B^{12}_2-B^{22}_1+A^{12}B^{12}-C^{22}B^{13}+C^{12}B^{23}+D^{12},\nonumber\\
 C^{14}&=&C^{12}_2-C^{22}_1+B^{12}C^{22}+A^{12}C^{12}-B^{22}C^{12}-C^{22}C^{13}+C^{12}C^{23}\nonumber\\
D^{14}&=&D^{12}_2-D^{22}_1+A^{12}D^{12}-B^{22}D^{12}+B^{12}D^{22}+C^{12}D^{23}-C^{22}D^{13},\nonumber\\
 A^{24}&=& A^{12}_1+B^{12}A^{12}+C^{12}A^{13}+D^{12},\quad  B^{24}=B^{12}_1+B^{12}B^{12}+C^{12}B^{13},\nonumber\\
C^{24}&=&C^{12}_1+B^{12}C^{12}+C^{12}C^{13},\quad D^{24}=B^{12}D^{12}+C^{12}D^{13}+D^{12}_1\\
 A^{34}&=& A^{13}_1+B^{13}A^{12}+C^{13}A^{13}+D^{13},\quad  B^{34}=B^{13}_1+B^{13}B^{12}+C^{13}B^{13},\nonumber\\
C^{34}&=&C^{13}_1+B^{13}C^{12}+C^{13}C^{13}, \quad D^{34}=D^{13}_1+B^{13}D^{12}+C^{13}D^{13}.\nonumber
\eea
Then  the integrability conditions for the system
\be\label{int21}
\partial_{x_j}{\bf w}={\bf A}^{(j )}{\bf w}\qquad  j=1,2,3,
\ee
must hold. They are
\be\label{int31}
{\bf A}^{(j)}_i-{\bf A}^{(i)}_j={\bf A}^{(i)}{\bf A}^{(j)}-{\bf A}^{(j)}{\bf A}^{(i)}\equiv [{\bf A}^{(i)},{\bf A}^{(j)}].
\ee
For convenience in the arguments to follow we set
\be\label{int3'1}
{\cal U}^{1}={\bf A}^{(3)}_2-{\bf A}^{(2)}_3- [{\bf A}^{(2)},{\bf A}^{(3)}],\ 
{\cal U}^{2}={\bf A}^{(1)}_3-{\bf A}^{(3)}_1- [{\bf A}^{(3)},{\bf A}^{(1)}], \ee
$${\cal U}^{3}={\bf A}^{(2)}_1-{\bf A}^{(1)}_2- [{\bf A}^{(1)},{\bf A}^{(2)}],
$$
so that the identities are
\be \label{int3''1} {\cal U}^1= {\cal U}^2={\cal U}^3=0.
\ee
The simplest of these, i.e., those that don't involve derivatives, are
\bea
A^{23}&=&B^{13}=C^{12},\quad B^{12}-A^{22}+A^{33}=C^{13}\nonumber\\
B^{23}-A^{13}&=&C^{22},\quad A^{12}+B^{33}=C^{23}.\label{int11}
\eea
Thus we can write all of the $A^{ij},B^{ij}, C^{ij}$ in terms of the 10 functions \be\label{10amigos}A^{12},A^{13},A^{22},A^{23},A^{33},B^{12},B^{22},B^{33},C^{33}.\ee
Also, identities (\ref{int3''1}) enable us to express each of $D^{33},D^{23},D^{22},D^{13},D^{12}$ as a polynomial in the $A^{ij},B^{ij},C^{ij}$ and their first derivatives for $i,j\le 3$.

As an example, the nondegenerate potential (\ref{nondegenpotential}) satisfies the canonical equations (\ref{veqn1a})
with
\be\label{nondegcanon} A^{33}=\frac{3}{x}+\frac{12x(z^2-x^2)}{1-r^4},\ B^{33}=\frac{12y(z^2-x^2)}{1-r^4},\ C^{33}=-\frac{3}{z} +\frac{12z(z^2-x^2)}{1-r^4},\ee
$$  D^{33}=\frac{24(z^2-x^2)}{1-r^4},
A^{22}=\frac{3}{x}+\frac{12x(y^2-x^2)}{1-r^4},\ B^{22}=-\frac{3}{y}+\frac{12y(y^2-x^2)}{1-r^4},$$
$$ C^{22}= \frac{12z(y^2-x^2)}{1-r^4},\   D^{22}=\frac{24(y^2-x^2)}{1-r^4},$$
$$A^{23}=\frac{12xyz}{1-r^4},\ B^{23}=\frac{12y^2z}{1-r^4},\ C^{23}=\frac{12yz^2}{1-r^4},\ D^{23}=\frac{24yz}{1-r^4},$$
$$A^{13}=\frac{12x^2z}{1-r^4},\ B^{13}=\frac{12xyz}{1-r^4},\ C^{13}=\frac{12xz^2}{1-r^4},\ D^{13}=\frac{24xz}{1-r^4},$$
$$A^{12}=\frac{12x^2y}{1-r^4},\ B^{12}=\frac{12xy^2}{1-r^4},\ C^{12}=\frac{12xyz}{1-r^4},\ D^{12}=\frac{24xy}{1-r^4},$$
where $r^2=x^2+y^2+z^2$. The nondegenerate potential (\ref{nondegenpotential1}) satisfies a simialr set of canonical equations but the expressions for the terms analogous to (\ref{nondegcanon}) are somwhat more complicated and would take two pages to list.

\subsection{Integrability conditions for the symmetries}
Since (as we assume)
the potential is nondegenerate, at any regular  point
${\bf x}_0$, $V$ and  the first derivatives $V_1,V_2,V_3$ can be chosen
arbitrarily. Thus the coefficients of $V$ and $V_j$ on both sides of 
equation (\ref{fundeqns12})  must be equal. From this, we obtain the relations
$$ 2a^{13}_{11}=a^{13}D^{33}+(a^{11}-a^{33})D^{13}+a^{12}D^{23}-a^{23}D^{12},
$$
$$
-3a^{31}_1=-a^{12}A^{23}+(a^{33}-a^{11})A^{13}+a^{23}A^{12}-a^{13}A^{33},
$$
$$
a^{12}_3-a^{32}_1=-a^{12}B^{23}+(a^{33}-a^{11})B^{13}+a^{23}B^{12}-a^{13}B^{33},
$$
$$
3a^{13}_3=-a^{12}C^{23}+(a^{33}-a^{11})C^{13}+a^{23}C^{12}-a^{13}C^{33},
$$
with 8 analogous relations from the other two Bertrand-Darboux
equations.
Using these 12 relations and equations   (\ref{symalgc}) we can solve for
all of the first partial derivatives $a^{jk}_i$ for $j\ne k$ and $a_\ell^{ii}-a^{jj}_\ell$ to obtain
\bea \label{symmetryeqnsc}
3a^{12}_1&=&a^{12}A^{22}-(a^{22}-a^{11})A^{12}-a^{23}A^{13}+a^{13}A^{23},\\
3(a^{11}-a^{22})_2&=&2\left[-a^{12}A^{22}+(a^{22}-a^{11})A^{12}+a^{23}A^{13}-a^{13}A^{23}\right],\nonumber\\
3a^{13}_3&=&-a^{12}C^{23}+(a^{33}-a^{11})C^{13}+a^{23}C^{12}-a^{13}C^{33}\nonumber\\
3(a^{33}-a^{11})_1&=&2\left[a^{12}C^{23}-(a^{33}-a^{11})C^{13}-a^{23}C^{12}+a^{13}C^{33}\right],\nonumber\\
3a^{23}_2&=&a^{23}(B^{33}-B^{22})-(a^{33}-a^{22})B^{23}-a^{13}B^{12}+a^{12}B^{13},\nonumber\\
3(a^{22}-a^{11})_3&=&2\left[-a^{23}(A^{12}+B^{33}-B^{22})+(a^{33}-a^{22})B^{23}+a^{13}(B^{12}+A^{33})\right.\nonumber\\
&&\left. +a^{12}(A^{23}-B^{13})+(a^{11}-a^{33})A^{13}\right],\nonumber\\
3a^{13}_1&=&-a^{23}A^{12}+(a^{11}-a^{33})A^{13}+a^{13}A^{33}+a^{12}A^{23},\nonumber\\
3(a^{33}-a^{11})_3&=&2\left[-a^{23}A^{12}+(a^{11}-a^{33})A^{13}+a^{13}A^{33}+a^{12}A^{23}\right],\nonumber\\
3(a^{33}-a^{11})_2&=&2\left[a^{13}(A^{23}-C^{12})+(a^{22}-a^{33})C^{23}+(a^{11}-a^{22})A^{12}\right.\nonumber\\
 &&\left. +a^{12}(A^{22}+C^{13})-a^{23}(A^{13}+C^{22}-C^{33})\right],\nonumber\\
3a^{23}_3&=&a^{13}C^{12}-(a^{22}-a^{33})C^{23}-a^{12}C^{13}-a^{23}(C^{33}-C^{22}),\nonumber\\
3a^{12}_2&=&-a^{13}B^{23}+(a^{22}-a^{11})B^{12}-a^{12}B^{22}+a^{23}B^{13},
\nonumber\\
3(a^{22}-a^{11})_1&=&2\left[a^{13}B^{23}-(a^{22}-a^{11})B^{12}+a^{12}B^{22}-a^{23}B^{13}\right],\nonumber\\
3a^{23}_1&=&a^{12}(B^{23}+C^{22})+a^{11}(B^{13}+C^{12})-a^{22}C^{12}-a^{33}B^{13}\nonumber\\
&+&a^{13}(B^{33}+C^{23})-a^{23}(C^{13}+B^{12}),\nonumber\\
3a^{12}_3&=&a^{12}(-2B^{23}+C^{22})+a^{11}(C^{12}-2B^{13})-a^{22}C^{12}+2a^{33}B^{13}\nonumber\\
&+&a^{13}(-2B^{33}+C^{23})+a^{23}(-C^{13}+2B^{12}),\nonumber\\
3a^{13}_2&=&a^{12}(B^{23}-2C^{22})+a^{11}(B^{13}-2C^{12})+2a^{22}C^{12}-a^{33}B^{13}\nonumber\\
&+&a^{13}(B^{33}-2C^{23})+a^{23}(2C^{13}-B^{12}).\nonumber
\eea
There are several conditions left over. These are the obstructions 
\be\label{obstruction1} a^{12}(C^{22}-B^{23}+A^{13})+(a^{11}-a^{22})(C^{12}-A^{23}) +(a^{11}-a^{33})(A^{23}-B^{13})\ee
$$ +a^{13}(C^{23}-B^{23}-A^{12})+a^{23}(B^{12}-C^{13}+A^{33}-A^{22})=0
$$
and 
\bea\label{obstruction2} 2a^{12}_{11}&=&a^{12}D^{22}+(a^{11}-a^{22})D^{12}+a^{13}D^{23}-a^{23}D^{13},\\
\label{obstruction3} 2a^{13}_{11}&=&a^{13}D^{33}+(a^{11}-a^{33})D^{13}+a^{12}D^{23}-a^{23}D^{12},\\
\label{obstruction4} 2a^{23}_{22}&=&a^{23}(D^{33}-D^{22})+(a^{22}-a^{33})D^{23}+a^{12}D^{13}-a^{13}D^{12}.\eea
It follows directly from conditions (\ref{int11}) for a nondegenerate potential that obstruction (\ref{obstruction1}) is satisfied identically.

\subsection{Nondegenerate potentials with 6 linearly independent second order conformal symmetries}
Suppose we have a superintegrable system with nondegenerate potential and 5 functionally independent second order symmetries ${\cal S}_1,\cdots, {\cal S}_4, {\cal H}$. We already know that any second order superintegrable Helmholtz system with nondegenerate potential on any conformally flat space will lead to such a Laplace system, in fact a system with 6 linearly independent second order symmetries. Now we will demonstrate that, conversely, every Laplace system with nondegenerate potential leads to a St\"ackel equivalence class of Helmholtz superintegrable systems on conformally flat manifolds. Let 
$${\cal H}\equiv p_1^2+p_2^2+p_3^2+V({\bf a},{\bf x})=0$$ be
a nondegenerate Laplace system, where the parameters are denoted by the vector ${\bf a}=(a_1,\cdots,a_5)$. Let $
\lambda({\bf x})\equiv V({\bf a}_0,{\bf x})$ be a special case of this potential with fixed parameters. With a suitable linear transformation in parameter space we can always assume
\be\label{specialpotential} V({\bf a},{\bf x})=\lambda({\bf x})v({\bf e},{\bf x})-a_5 \lambda({\bf x}), \quad {\bf a}=({\bf e},a_5).\ee
By assumption, both $V$ and its special case $\lambda$ satisfy the canonical equations (\ref{veqn1a}) for the potential. Substituting $V=\lambda U$ in these equations we see that the St\"ackel transformed potential $U$ 
satisfies the canonical equations 
\be\label{veqn2a}\ba{lllll}
U_{22}&=&U_{11}&+&{\tilde A}^{22}U_1+{\tilde B}^{22}U_2+{\tilde C}^{22}U_3,\\
U_{33}&=&U_{11}&+&{\tilde A}^{33}U_1+{\tilde B}^{33}U_2+{\tilde C}^{33}U_3,\\
U_{12}&=& &&{\tilde A}^{12}U_1+{\tilde B}^{12}U_2+{\tilde C}^{12}U_3,\\
U_{13}&=& &&{\tilde A}^{13}U_1+{\tilde B}^{13}U_2+{\tilde C}^{13}U_3,\\
U_{23}&=& &&{\tilde A}^{23}U_1+{\tilde B}^{23}U_2+{\tilde C}^{23}U_3,\ea
\ee
characteristic of a Helmholtz system with nondegenerate potential, where
$$ {\tilde A}^{33}=A^{33}+2\frac{\lambda_1}{\lambda},\ {\tilde B}^{33}=B^{33},\ {\tilde C}^{33}=C^{33}-2\frac{\lambda_3}{\lambda},\  {\tilde A}^{22}=A^{22}+2\frac{\lambda_1}{\lambda},$$
$$ {\tilde B}^{22}=B^{22}-2\frac{\lambda_2}{\lambda},\ {\tilde C}^{22}=C^{22},
{\tilde A}^{12}=A^{12}-\frac{\lambda_2}{\lambda},\ {\tilde B}^{12}=B^{12}-\frac{\lambda_1}{\lambda},\ {\tilde C}^{12}=C^{12},$$
$$ {\tilde A}^{13}=A^{13}-\frac{\lambda_3}{\lambda},\ {\tilde B}^{13}=B^{13},\ {\tilde C}^{13}=C^{13}-\frac{\lambda_1}{\lambda},$$
$${\tilde A}^{23}=A^{23},\ {\tilde B}^{23}=B^{23}-\frac{\lambda_3}{\lambda},\ {\tilde C}^{23}=C^{23}-\frac{\lambda_2}{\lambda}.$$
In particular, the coefficients of $U$ vanish in all these equations. The corresponding St\"ackel transformed system is
\be\label{Staceltrans} {\tilde {\cal H}}\equiv \frac{\cal H}{\lambda}\equiv \frac{p_1^2+p_2^2+p_3^2}{\lambda}+U\equiv
\frac{p_1^2+p_2^2+p_3^2}{\lambda}+v-a_5=0.\ee
This appears to be a nondegenerate Helmholtz superintegrable system with energy $a_5$. However, we know only that ${\cal S}_1,\cdots, {\cal S}_5$ are conformal symmetries of this system, not the required true symmetries. We need to exhibit true symmetries. Recalling the conclusion of Theorem \ref{stackelt}, if $\cal S $ is a second order conformal symmetry of ${\cal H}=0$ then ${\tilde{\cal S} }={\cal S}-\frac{W_\lambda}{\lambda}{\cal H}$ is a true symmetry of (\ref{Staceltrans}). 
Thus the new conformal symmetries 
$${\tilde {\cal S}}_h={\cal S}_h-\frac{W_\lambda}{\lambda}{ {\cal H}},\quad h=1,\cdots,5,$$
are actually true symmetries, i.e.,  in involution with the Hamiltonian 
$(p_1^2+p_2^2+p_3^2)/\lambda+v({\bf e},{\bf x})$.  Note that the symmetries ${\tilde {\cal S}}_h$, ${\cal S}_h$ agree on the hypersurface ${\tilde {\cal H}}=0$ and $\{{\tilde {\cal S}}_h,{\tilde {\cal S}}_t\}$ vanishes on the hypersurface if and only if $\{ {\cal S}_h,{\cal S}_t\}$ vanishes. 

We see that a Laplace superintegrable system with nondegenerate potential and a guaranteed 5 functionally independent second order conformal symmetries is equivalent via a St\"ackel transform to a Helmholtz system with nondegenerate potential and 5 functionally independent second order true symmetries. However the $5\Longrightarrow 6$ Theorem in  \cite{KKM20051} shows that such a Helmholtz system actually admits 6 linearly independent true symmetries. This extra symmetry must correspond to a conformal symmetry of the original Laplace system. Thus the Laplace system must admit 5 conformal symmetries, in addition to the Hamiltonian.
\begin{theorem}\label{4implies5} $\left(4\Longrightarrow 5\right)$  A Laplace second order superintegrable system with functionally independent generators    ${\cal S}_1,\cdots, {\cal S}_4, {\cal H}$, and nondegenerate potential admits a 5th conformal second order symmetry ${\cal S}_5$ such that the set  $\{{\cal S}_1,\cdots,  {\cal S}_5\}$ is linearly independent on the hypersurface ${\cal H}=0$. \end{theorem}
\begin{theorem} There is a one-to-one relationship between 
 flat space  Laplace systems with nondegenerate potential   and St\"ackel equivalence classes of superintegrable Helmholtz systems with nondegenerate potential on  conformally flat spaces.
 \end{theorem}
For such a Laplace system
the integrability conditions for the potential (\ref{int3''1}) and the integrability conditions for the symmetries (\ref{symmetryeqnsc}), some 150 equations in all, are satisfied identically. Furthermore each of the obstruction equations (\ref{obstruction2}), (\ref{obstruction3}), (\ref{obstruction4}) must also be satisfied identically.

By a straightforward but lengthy calculation with MAPLE we  established the following from these equations:
\begin{enumerate} \item We found 30 equations expressing the 30 partial derivatives $\partial_iF^{jk}$ as quadratic polynomials in the 10 functions $A^{12},A^{13},A^{22},A^{23},A^{33}$, $B^{12}$, $B^{22}$, $B^{23},B^{33},C^{33}$. Here $F^{jk}$ is any one of these functions.
For example, three of these equations are
\bea
\partial_xA^{12}&=& \frac13 A^{23}A^{13}+A^{23}B^{23}+B^{33}A^{33}-\frac13 A^{12}A^{22}-\frac13 A^{12}B^{12}\nonumber\\
&-&B^{33}A^{22}-A^{23}C^{33},
\nonumber\\
\partial_y A^{12}&=&\frac12 (A^{23})^2+\frac16 (A^{12})^2+\frac12(B^{12})^2+\frac16(A^{33})^2-\frac16 C^{33}A^{13}-\frac16 (B^{33})^2 \nonumber\\
&-&\frac16 A^{22}A^{33}-\frac13 B^{33}A^{12}+\frac13 A^{33}B^{12}+\frac16 B^{33}B^{22}-\frac16 (B^{23})^2+\frac16 C^{33}B^{23},
\nonumber\\
\partial_z A^{12}&=&\frac13 A^{23}A^{33}+\frac23 B^{12}A^{23}+\frac13 A^{13}A^{12}-\frac13 A^{23}A^{22}. 
\nonumber\eea
\item We found quadratic expressions for each of the terms  $D^{ij}$. Indeed:
 \bea\label{Dexpressions} D^{12}&=&\frac23(- A^{12} B^{12}+ A^{23} B^{23}+B^{33} A^{33}- B^{33}A^{22}-A^{23} C^{33}),\\
 D^{13}&=&\frac23 (- A^{13}B^{12}- B^{22}A^{23}-B^{23}A^{33}+ B^{23} A^{22}+A^{23} B^{33}+A^{12} A^{23}),\nonumber\\
 D^{22}&=&\frac23\left((A^{12})^2+ B^{12}A^{22}- B^{22} A^{12}+ C^{33} A^{13}+2 B^{33} A^{12}-2 A^{33} B^{12}- C^{33} B^{23}\right.\nonumber\\
 &-&\left. B^{33} B^{22}+(B^{23})^2+ (B^{33})^2- (B^{12})^2- (A^{33})^2+ A^{33} A^{22}-B^{23} A^{13}\right),\nonumber\\
 D^{23}&=&\frac23( A^{23}A^{33}+B^{12}A^{23}-B^{23} A^{12}- A^{13} B^{33}),\nonumber\\
 D^{33}&=&\frac23(B^{33} A^{12}- (B^{12})^2- A^{33} B^{12}-C^{33} B^{23}- B^{33} B^{22}+ (B^{23})^2+(B^{33})^2).\nonumber\eea
\item There are no other consequences of these identities.
\end{enumerate}

There are no polynomial identities that the 10 functions must obey. What is amazing is that the  integrability conditions for the expressions $\partial_i F^{jk}$ are satisfied identically! 
\begin{theorem} \label{basicresult}Suppose the integrability conditions (\ref{int3''1}) and the integrability conditions for the symmetries (\ref{symmetryeqnsc}), are satisfied identically. Then equations (\ref{Dexpressions}) hold and the integrability conditions $\partial_\ell(\partial_iF^{jk})=\partial_i(\partial_\ell F^{jk})$ are satisfied identically. Thus at a regular point ${\bf x}_0$ we can choose a 10-tuple ${\bf c}=(c_1,\cdots,c_{10})$ arbitrarily and there will exist one and only one superintegrable system such that
$$ \left (A^{12}({\bf x}_0),A^{13}({\bf x}_0),\cdots, B^{33}({\bf x}_0),C^{33}({\bf x}_0)\right)={\bf c}.$$ 
\end{theorem}
\subsection{The classification problem for nondegenerate potentials}
Theorem \ref{basicresult} provides the basis for classifying all manifolds for which Helmholtz superintegrable systems exist, and all nondegenerate potentials on these manifolds, a program that is not yet complete. Indeed, note that 
the conformal group, with connected component isomorphic to  $SO(5,{\bb C})$,  acts naturally on the 10-tuple  $\bf c$. Suppose we have a conformal superintegrable system
\be\label{initialsystem}p_1^2+p_2^2+p_3^2+V({\bf a},{\bf x})=0.\ee
If $g:{\bf x}\to {\bf x}'={\bf x}g$ is an element of the conformal group (considered as a transformation group) then $g$ acts on functions   $f({\bf x})$ via operators $T(g)$ such that  $T(g)f ({\bf x})=f({\bf x}g)$. Then $T(g_1g_2)=T(g_1)T(g_2)$ so we get a representation. Under this action $p_1^2+p_2^2+p_3^2\to {p'_1}^2+{p'_2}^2+{p'_3}^2=c({\bf x},g)({p_1}^2+{p_2}^2+{p_3}^2)$ where $c({\bf x},g)$ is the conformality factor. Thus the conformally superintegrable system transforms to another conformally superintegrable system
\be\label{targetsystem} {p_1}^2+{p_2}^2+{p_3}^2+V'({\bf x})=0,\quad V'({\bf x})=\frac{1}{c({\bf x},g)}  V({\bf a},{\bf x}g).\ee
The conformal symmetries transform in an obvious manner. System (\ref{initialsystem}) is uniquely characterized by the values of the 10-tuple ${\bf c}_0$ at the regular point ${\bf x}_0$.  There is an induced action $g:{\bf c}\to {\bf c}'={\bf c}g$ of the conformal group on 10-tuples such that system (\ref{targetsystem}) is uniquely determined by the values ${\bf c}_0g$ at the point ${\bf x}_0g$.
Since the equations determining this action locally on 10-tuples are autonomous for Euclidean actions, we can mostly ignore the starting point ${\bf x}_0$ and focus just on the map ${\bf c}\to{\bf c}g$. However, for general conformal actions we have to consider the map on the $13$-dimensional manifold of points $({\bf x},{\bf c})$. Thus $({\bf x}_0,{\bf c})\to ({\bf x}_0 g,{\bf c}g)$. Clearly, all conformally superintegrable systems  related by elements of the conformal group have essentially the same structure and should be identified. 
Thus we say  that two superintegrable systems are equivalent if and only if they are on the same orbit under the action  $({\bf x},{\bf c})\to ({\bf x}g,{\bf c}g)$.
It appears that we might be able to determine a solution in each equivalence class and thus, indirectly, find all 3D nondegenerate Helmholtz superintegrable systems, including all those on nonflat spaces.

To see how the classification might proceed, recall from \cite{KKM20051} that every Helmholtz superintegrable system with nondegenerate potential is St\"ackel equivalent to a system on either flat space, or the complex 3-sphere, or both. Thus to classify all possible systems we need only to classify the constant curvature space systems. From the point of view of this paper, the flat space systems are just those for which $D^{ij}\equiv 0$ in equations (\ref{Dexpressions}). Thus $I^{(a)}=\cdots =I^{(e)}=0$ at each regular point, where
\begin{eqnarray}\label{ideal}
I^{(a)} &=& -A^{22}B^{23} + B^{23}A^{33} + B^{12}A^{13}
     + A^{23}B^{22} - A^{12}A^{23} - A^{23}B^{33} \\
I^{(b)} &=& (A^{33})^2 + B^{12}A^{33} - A^{33}A^{22}
     - A^{12}B^{33} - A^{13}C^{33} + A^{12}B^{22}\nonumber \\
& & {}\qquad
     - B^{12}A^{22} + A^{13}B^{23} - (A^{12})^2 \nonumber\\
I^{(c)} &=& B^{23}C^{33} + B^{12}A^{33} + (B^{12})^2
     + B^{22}B^{33} - (B^{33})^2 - A^{12}B^{33} - (B^{23})^2\nonumber \\
I^{(d)} &=& -B^{12}A^{23} - A^{33}A^{23} + A^{13}B^{33}
            + A^{12}B^{23}\nonumber \\
I^{(e)} &=& -B^{23}A^{23} + C^{33}A^{23} + A^{22}B^{33}
            - A^{33}B^{33} + B^{12}A^{12}. \nonumber
\end{eqnarray}
In \cite{KKM20071} it was  shown via  Gr\"obner basis methods that  these conditions further imply  $I^{(f)}=0$ where
\begin{eqnarray}\label{ideal2} I^{(f)} &=& A^{13}C^{33} + 2A^{13}B^{23} + B^{22}B^{33}
     - (B^{33})^2 + A^{33}A^{22} - (A^{33})^2\\
& & {}\quad + 2A^{12}B^{22} 
     + (A^{12})^2 - 2B^{12}A^{22} + (B^{12})^2
     + B^{23}C^{33} - (B^{23})^2 - 3(A^{23})^2, \nonumber
\end{eqnarray}
and that any system satisfying  conditions $I^{(a)}=\cdots =I^{(f)}=0$ at some regular point uniquely defines a flat space superintegrable system. From this we were able to classify all flat space Helmholtz superintegrable systems. To compare our present notation to the results of  \cite{KKM20051},  consider Laplace superintegrable systems with nondegenerate potential that are St\"ackel equivalent to a Helmholtz superintegrable system on the complex 3-sphere.  It is always possible to choose coordinates  $x,y,z$ on the 3-sphere such that the metric takes the form $\lambda(x,y,z)=4/(1+x^2+y^2+z^2)^2$ as in (\ref{metric}).  Thus the nondegenerate potential for the associated flat space Laplace system must have  $V=\lambda(x,y,z)$ as a particular instance. Substituting this requirement in equations (\ref{veqn1a}) we can solve for the functions $D^{ij}$ to get
\begin{eqnarray}\label{newDterms} D^{12}&=& -\frac23 I^{(e)}=G_{xy}+G_xG_y-A^{12}G_x-B^{12}G_y-C^{12}G_z,\\
 D^{13}&=& -\frac23 I^{(a)}=G_{xz}+G_xG_z-A^{13}G_x-B^{13}G_y-C^{13}G_z,\nonumber\\
D^{22}&=& -\frac23 I^{(a)}+\frac23 I^{(b)}=G_{yy}+G_y^2-G_{xx}-G_x^2-A^{22}G_x-B^{22}G_y-C^{22}G_z,\nonumber\\
 D^{23}&=& -\frac23 I^{(d)}=G_{yz}+G_yG_z-A^{23}G_x-B^{23}G_y-C^{23}G_z,\nonumber\\
D^{33}&=& -\frac23 I^{(a)}=G_{zz}+G_z^2-G_{xx}-G_x^2-A^{33}G_x-B^{33}G_y-C^{33}G_z,\nonumber
\end{eqnarray}
where $G(x,y,z)=\ln\lambda$. Equations (\ref{Dexpressions}) and (\ref{newDterms}) should play an important role in the classification of all Helmholtz superintegrable systems on the complex 3-sphere with nondegenerate potential.

\end{document}